\documentclass[12pt]{article}

\font\grande=cmr9.5 scaled \magstep4
\font\medio=cmr9.5 scaled \magstep2
\outer\def\beginsection#1\par{\medbreak\bigskip
      \message{#1}\leftline{\bf#1}\nobreak\medskip
\vskip-\parskip
      \noindent}
\usepackage{graphicx} 
\begin{document}
\bibliographystyle {unsrt}

\titlepage

\begin{flushright}
CERN-PH-TH-2015-059
\end{flushright}

\vspace{15mm}
\begin{center}
{\grande Viscous modes, isocurvature perturbations}\\
\vspace{5mm}
{\grande and CMB initial conditions}\\
\vspace{15mm}
 Massimo Giovannini 
 \footnote{Electronic address: massimo.giovannini@cern.ch} \\
\vspace{0.5cm}
{{\sl Department of Physics, Theory Division, CERN, 1211 Geneva 23, Switzerland }}\\
\vspace{1cm}
{{\sl INFN, Section of Milan-Bicocca, 20126 Milan, Italy}}
\vspace*{1cm}

\end{center}

\vskip 1cm
\centerline{\medio  Abstract}
\vskip 1cm
When the predecoupling plasma is thermodynamically reversible its fluctuations 
are classified in terms of the adiabatic and entropic modes.  
A different category of physical solutions, so far unexplored, arises when the  
inhomogeneities of the viscosity coefficients induce computable curvature perturbations. 
The viscous modes are explicitly illustrated and compared with the conventional isocurvature solutions. 
\noindent

\vspace{5mm}

\vfill
\newpage
The adiabatic and the entropic modes have been introduced long ago \cite{peebles} (see also \cite{hh} and references therein) in connection with the pioneering analyses of the temperature and polarization anisotropies of the Cosmic Microwave Background (CMB in what follows). The relative 
position of the first anticorrelation peak of the cross-spectrum between the temperature and the polarization
demonstrated (already from the first releases of the WMAP data \cite{WMAP1})
that the initial conditions of the Einstein-Boltzmann hierarchy are predominantly 
adiabatic. This evidence has been further scrutinized and quantitatively refined by the subsequent WMAP releases and by Planck explorer results \cite{WMAP2}. While the conventional distinction between the adiabatic and the entropic solutions 
posits that the ambient fluid is thermodynamically reversible, it seems plausible to relax this hypothesis 
 with the aim of complementing and extending the physical initial conditions of the predecoupling plasma. 
According to the current data \cite{WMAP1,WMAP2}, anticorrelated non-adiabatic modes in the presence of a dominant adiabatic mode may even improve 
the fit of the temperature autocorrelations accounting for potential large-scale suppressions of the corresponding angular power spectra. This same strategy could also be investigated in the case of the viscous mode derived in this analysis. 
 
The fluctuations of the total pressure of the predecoupling plasma are customarily decomposed into an adiabatic component supplemented by the entropic (or simply non-adiabatic) contributions:
\begin{equation}
\delta p_{t} = c_{\mathrm{st}}^2 \delta \rho_{t} +  \delta p_{\mathrm{nad}}, \qquad c_{\mathrm{st}}^2 =\biggl(\frac{\delta p_{t}}{\delta\rho_{t}}\biggr)_{\varsigma},\qquad \delta p_{\mathrm{nad}}= \biggl( \frac{\delta p_{t}}{\delta \varsigma }\biggr)_{\rho_{t}} \delta\varsigma,
\label{dpnad}
\end{equation}
where the subscripts in the brackets signify that the variations must be performed, respectively, at constant specific entropy $\varsigma$ and constant  energy density $\rho_{t}$. If the cold dark matter (CDM in what follows) is only supplemented by 
a radiation background, the specific entropy is the ratio between the entropy density of the radiation and the CDM concentration, i.e. approximately $\varsigma \simeq T^3/n_{\rm c}$. The entropy contrast 
${\mathcal S} = \delta\varsigma/\varsigma$ can then be written as ${\mathcal S}= (3 \delta_{r}/4 - \delta_{c})$ where $\delta_{r}$ and $\delta_{c}$ are, respectively, the density contrasts of the radiation and of the CDM species. If ${\mathcal S} =0$ the density contrasts 
are in a fixed ratio (i.e. $\delta_{r}/ \delta_{c} = 4/3$) and the solution is adiabatic;
if ${\mathcal S} \neq 0$ the adiabaticity condition is violated and $\delta_{r} \neq 4 \delta_{c}/3$ (see e.g. \cite{hh}).
Across the matter-radiation transition the entropic solutions are determined by the total sound speed of the plasma\footnote{We shall consider hereafter the case 
of a conformally flat background geometry characterized by a scale factor $a(\tau)$ where $\tau$ denotes the conformal time coordinate. } and by the 
specific form of $\delta p_{\mathrm{nad}}$; for the CDM-radiation mode we have:
\begin{equation}
c_{\mathrm{st}}^2 = \frac{4}{3( 3 \alpha + 4)}, \qquad \delta p_{\mathrm{nad}} = \rho_{\mathrm{c}} c_{\mathrm{st}}^2 {\mathcal S},
\label{cs2}
\end{equation}
where $\rho_{\mathrm{c}}$ denotes the CDM energy density $\alpha = a/a_{\mathrm{eq}}$ is the 
scale factor normalized at the matter-radiation equality. The analytic expression of $\delta p_{\mathrm{nad}}$ 
can be generalized to a generic pair of species of the plasma (see e.g. fourth article in Ref. \cite{hh}).

Even if the conventional terminology might suggest otherwise, the non-adiabatic modes arise in a globally inviscid fluid, as the preceding considerations 
illustrate. In the present paper we want to drop this hypothesis since the total energy-momentum tensor of the plasma could include the contributions of the shear viscosity, of the bulk viscosity and of the heat transfer. The adiabatic limit, in a strict sense, is recovered when the viscous contributions are neglected and the total entropy four-vector is conserved. 
Across the matter-radiation transition the shear viscosity coefficient $\eta$  determines the optical depth, the Silk damping scale and, ultimately, the shape of the 
visibility function \cite{peebles,visibility}. Conversely the role of the bulk viscosity coefficient $\xi$ is disregarded since it must vanish in the limit $c_{\mathrm{st}} \to 1/\sqrt{3}$ \cite{BV1}.  The scaling properties of the viscous coefficient can be expressed, in a unified notation, as: 
\begin{equation}
\frac{\xi}{\eta} \simeq \frac{\xi}{ T \chi} \simeq \biggl( \frac{1}{3}- c_{\mathrm{st}}^2 \biggr)^{2 q}, \qquad \frac{\eta}{\chi} \simeq T,
\label{scaling}
\end{equation}
where $\chi$ denotes the coefficient of heat conduction; the values of $q\geq 1$ parametrize the rate of extinction of the bulk viscosity coefficient prior to matter-radiation equality, i.e. in the limit  $\tau \ll \tau_{\mathrm{eq}}$. It is possible to argue \cite{BV1}, with reasonable arguments, that $q\simeq 1$
in Eq. (\ref{scaling}); however, to avoid supplementary hypotheses, the values $q$ will be taken as a free parameter.

Prior to matter radiation equality but well below the temperature of neutrino decoupling, the total energy-momentum tensor of the fluid shall then be written, for the present ends, as:
\begin{equation}
{\mathcal T}_{\mu}^{\nu}(\rho_{t},\,n_{t},\, \xi) = (p_{\mathrm{t}} + \rho_{\mathrm{t}}) u_{\mu} u^{\nu} - p_{\mathrm{t}} \delta_{\mu}^{\nu} + \xi\,\biggl( \delta_{\mu}^{\nu} - u_{\mu} u^{\nu} \biggr) \,\nabla_{\alpha} u^{\alpha}.
\label{BVC1}
\end{equation}
The total pressure of the plasma $p_{t}$ is in general defined to be the same 
function of $\rho_{t}$ (i.e. the total energy density) and $n_{t}$ (i.e. the total concentration) as 
in the adiabatic limit. The Eckart approach\footnote{In the Landau approach the four-velocity coincides with the velocity of the energy transport while in the framework of the Eckart theory 
 $u_{\mu}$ denotes the velocity of the particle transport \cite{LL}.} 
 seems preferable in the present situation where the concentration of radiation quanta exceeds the concentrations of the other species. 
The first principle of thermodynamics together with the covariant conservation of the total energy momentum tensor and of the global particle current (i.e. $j^{\alpha} = n_{t}\, u^{\alpha}$) implies the law of growth of the entropy and the rate of suppression of the temperature:
\begin{equation}
\nabla_{\mu} S^{\mu} = \frac{\xi}{T} (\nabla_{\alpha}u^{\alpha})^2 , \qquad u^{\mu} \nabla_{\mu} T = - T c_{\mathrm{st}}^2 \nabla_{\mu} u^{\mu},
\label{ENTR1}
\end{equation}
where $S^{\mu}=n_{t}\,S \,u^{\mu}$ is the entropy four-vector in the case when the bulk viscosity is the only source of irreversibility\footnote{Note that $T$ is the temperature of the plasma that coincides, roughly speaking, with the photon temperature since the concentration of the photons greatly exceeds the concentration of the other species both relativistic and non-relativistic.}.

The bulk viscosity coefficient $\xi$ is the sum of a homogeneous part supplemented by the inhomogeneous contribution:
\begin{equation}
\xi(\vec{x},\tau) = a(\tau) \overline{\xi}(\tau) +a(\tau)  \delta \xi(\vec{x},\tau).
\label{decomp}
\end{equation}
The evolution equations for the homogeneous expansion rate are a consequence of  Eq. (\ref{BVC1}) and of the Einstein equations:
\begin{equation}
 ({\mathcal H}^2 - {\mathcal H}') = 4 \pi G a^2 (\rho_{\mathrm{t}} + {\mathcal P}_{\mathrm{t}}), \qquad 3 {\mathcal H}^2 = 8 \pi G a^2 \rho_{\mathrm{t}},
 \label{FL}
 \end{equation}
 where ${\mathcal P}_{t}= p_{t} - 3 \overline{\xi} \, {\mathcal H}$  is the shifted background pressure. The gauge-invariant bulk viscosity 
fluctuations are given by\footnote{We decompose 
the scalar fluctuations of the four-dimensional metric tensor $ g_{\mu\nu}$ as $\delta  g_{0i} = - a^2  \partial_{i} B$, $\delta g_{ij} = 
2 a^2 (\psi \delta_{i j} - \partial_{i} \partial_{j} E)$ and $\delta g_{00} = 2 a^2 \phi$. 
The two gauge-invariant Bardeen combinations correspond to $\Phi = \phi + ( B - E')' + {\cal H} ( B - E')$ and  $\Psi= \psi - {\cal H} ( B - E')$.}:
\begin{equation}
\Xi =  \delta \xi + [\overline{\xi}' + {\mathcal H} \overline{\xi} ]( B - E'),
\label{XI}
\end{equation}
where $B$ and $E$ parametrize the off-diagonal (scalar) degrees of freedom of the perturbed metric. 
If $\overline{\xi}(\tau) =\overline{\xi}^{\prime}(\tau) = 0$  then $\delta \xi$ equals $\Xi$ and the bulk 
viscous fluctuations are automatically gauge-invariant; in the same interesting limit ${\mathcal P}_{t} \to p_{t}$ and the shifted 
pressure coincides with the pressure itself.
The Hamiltonian and the momentum constraints for the evolution of the fluctuations 
are not modified by $\Xi$. On the contrary $\Xi$ appears in the $(ij)$ components of the perturbed Einstein equations. 
Consequently the evolution of the curvature perturbations on comoving orthogonal hypersurfaces is given by:
\begin{equation}
 {\mathcal R}' - \frac{ 3 {\mathcal H}}{( \rho_{t} + {\mathcal P}_{t})} (\overline{\xi}' + {\mathcal H} \overline{\xi}) ( {\mathcal R} + \Psi) 
=  \frac{3 {\mathcal H}^2 }{ \rho_{t} + {\mathcal P}_{t}} \Xi 
+ \overline{\xi} \frac{{\mathcal H}}{\rho + {\mathcal P}_{t}} \Theta_{t} - \frac{{\mathcal H} c_{\mathrm{st}}^2}{4\pi G a^2(\rho_{t} + {\mathcal P}_{t})}
\nabla^2 \Psi,
\label{evolR}
\end{equation}
where  $\Phi$ and $\Psi$ denote the Bardeen potentials entering the definition of the curvature perturbations as  ${\mathcal R} = - [\Psi + {\mathcal H} ({\mathcal H} \Phi + \Psi^{\prime})/({\mathcal H}^2 - {\mathcal H}^{\prime})]$. In Eq. (\ref{evolR}) $\Theta_{t}= \vec{\nabla}\cdot \vec{v}_{t}$ is the three-divergence of the gauge-invariant velocity field. On the right-hand side of Eq. 
(\ref{evolR}) the only term surviving the large-scale limit is the first one; $\nabla^2 \Psi$ is subleading and, similarly, $\Theta_{t}$ can be neglected thanks to the momentum constraint
implying that 
\begin{equation}
\Theta_{t} = \frac{\nabla^2( \Psi^{\prime} + {\mathcal H} \Phi)}{({\mathcal H}^2 - {\mathcal H}^{\prime})} = - \frac{\nabla^2( {\mathcal R} + \Psi)}{{\mathcal H}},
\label{MC}
\end{equation}
where the second equality in Eq. (\ref{MC}) follows from the gauge-invariant definition of ${\mathcal R}$. Equation (\ref{MC}) shows, as it is well known, that the total 
velocity field can be neglected in the large-scale limit.

The total viscosity is given by $ \Xi(k,\tau) = \Xi_{\gamma}(k,\tau) +  \Xi_{\nu}(k,\tau)$ and it is proportional 
to the total radiation energy density $\rho_{R} = \rho_{\gamma}+ \rho_{\nu}$; note that  $\Xi_{\gamma}(k,\tau)$ and $ \Xi_{\nu}(k,\tau)$ are 
the viscosities of the photon and of the neutrino sectors. 
According to Eq. (\ref{scaling}) $\Xi_{\gamma}$ and $\Xi_{\nu}$ go to zero deep in the radiation epoch and they must be proportional 
to the corresponding energy densities; thus they can be parametrized, in Fourier space, as:
\begin{equation}
\Xi_{\gamma}(k,\tau) = \frac{\rho_{\gamma}}{{\mathcal H}} \biggl[ \frac{1}{3}-c_{\mathrm{st}}^2(\tau)  \biggr]^{2 q} {\mathcal V}_{*}(k), 
\qquad \Xi_{\nu}(k,\tau) = \frac{\rho_{\nu}}{{\mathcal H}}\biggl[ \frac{1}{3}-c_{\mathrm{st}}^2(\tau)  \biggr]^{2 q}{\mathcal V}_{*}(k),
 \label{vv1}
 \end{equation}
 where ${\mathcal V}_{*}(k)$ denotes the fully inhomogeneous contribution\footnote{ If the inhomogeneities are localized either in CDM or in the baryon sector Eq. (\ref{vv1}) can be complemented, respectively, by $\Xi_{c}$ and 
$\Xi_{b}$; hereafter we shall focus, for illustrative purposes, on the case where $\Xi_{b} = \Xi_{c} =0$. }.

Barring for a potentially important role of the bulk viscosity coefficient in the very early Universe, as repeatedly suggested in the past (see for instance \cite{BV2,BV3}), $\overline{\xi}$ cannot appreciably affect the evolution of the homogeneous expansion rate across the matter-radiation transition.  
We shall therefore accept that $\overline{\xi} = \overline{\xi}^{\prime}=0$, so that the solution of Eq. (\ref{FL}) across the matter-radiation transition will be given, in conformal time,  by $\alpha= a/a_{eq}=(x^2 + 2 x)$ where $ x = \tau/\tau_{1}$  and $\tau_{\mathrm{eq}} = (\sqrt{2} -1) \tau_{1}$.  
 Equation (\ref{evolR}) can then be solved in the large-scale limit using $\alpha$ as integration variable and recalling that, by definition, ${\mathcal H} = \partial \ln{\alpha}/\partial\alpha$; the result of this step can be written as:
\begin{equation}
{\mathcal R}(k,\alpha) = {\mathcal R}_{*}(k) +\frac{9}{8 q} {\mathcal V}_{*}(k) \frac{\alpha^{2 q}}{(3 \alpha + 4)^{2 q}}.
\label{solR}
\end{equation}
In Eq. (\ref{solR})  ${\mathcal R}_{*}(k)$ denotes the conventional adiabatic mode while ${\mathcal V}_{*}(k)$ 
corresponds to the viscous mode. Recalling the definition of ${\mathcal R}$ in terms of $\Psi$,  the expression for $\Psi(k,\alpha)$ can be derived in terms of hypergeometric functions and the two relevant asymptotic limits (well before and well after equality) are:
\begin{eqnarray}
\lim_{\alpha \ll 1} \Psi(k,\alpha) &\to& - \frac{2}{3} {\mathcal R}_{*}(k) \biggl[ 1 - \frac{\alpha}{16} + {\mathcal O}(\alpha^2)\biggr] - \frac{9}{4 q ( 2 q + 3)} {\mathcal V}_{*}(k) \biggl(\frac{\alpha}{4}\biggr)^{2 q},
 \nonumber\\
 \lim_{\alpha \gg 1} \Psi(k,\alpha) &\to&- \frac{3}{5} {\mathcal R}_{*}(k)\biggl[ 1 + \frac{2}{9\alpha} + {\mathcal O}(\alpha^{-2})\biggr]  - \frac{3^{3 - 2 q}}{40 q} {\mathcal V}_{*}(k).
 \label{psilim}
 \end{eqnarray}

The derivation of Eq. (\ref{psilim}) assumes that $\Phi = \Psi$ and implies the absence of any source of anisotropic stress; while this approximation holds over large-length scales, the presence of the neutrinos introduces a difference between $\Phi$ and $\Psi$. The full structure of the viscous mode will now be obtained in the limit $\alpha \ll 1$ and in the presence of the 
neutrinos, taken to be massless as in the context of the concordance paradigm. The gauge-invariant evolution equations for the lowest multipoles 
of the neutrino hierarchy in the presence of viscous inhomogeneities can be written as:
\begin{eqnarray}
&& \delta_{\nu}^{\prime} = 4 \Psi'  + \frac{9 {\mathcal H}^2}{\rho_{\nu}} \Xi_{\nu} - \frac{4}{3} \theta_{\nu},
\label{nu1}\\
&& \theta_{\nu}^{\prime} = \frac{k^2}{4} \delta_{\nu} - k^2 \sigma_{\nu} + k^2 \Phi - \frac{9 {\mathcal H}}{4 \rho_{\nu}} k^2 \Xi_{\nu},
\qquad \sigma_{\nu}^{\prime} = \frac{4}{15} \theta_{\nu}- \frac{3}{10} {\mathcal F}_{\nu\, 3},
\label{nu3}
\end{eqnarray}
where ${\mathcal F}_{\nu 3}$ is the octupole of the perturbed phase space distribution of the neutrinos. 
In the tight-coupling limit the evolution equations of the baryon-photon system are instead:
\begin{eqnarray}
&& \delta_{\gamma}^{\prime} = 4 \Psi' + \frac{9 {\mathcal H}^2}{\rho_{\gamma}} \Xi_{\gamma} - \frac{4}{3} \theta_{\gamma\, b},
\qquad \delta_{b}^{\prime} = 3 \Psi' - \theta_{\gamma\,b} + \frac{9{\mathcal H}^2}{\rho_{b}} \Xi_{b},
\label{pb2}\\
&& \theta_{\gamma\,b}^{\prime} + \frac{{\mathcal H} R_{b}}{R_{b} + 1} \theta_{\gamma\, b} + \frac{\eta}{\rho_{\gamma}} \frac{k^2 \theta_{\gamma\, b}}{(R_{b} +1)} 
= k^2 \Phi + \frac{k^2 \delta_{\gamma}}{4 ( R_{b} +1)} - \frac{9}{4} \frac{{\mathcal H}}{\rho_{b}} k^2 ( \Xi_{b} + \Xi_{\gamma}).
\label{pb3}
\end{eqnarray}
In Eq. (\ref{pb3}), as usual,  $R_{b} =  3 \rho_{b}/(4 \rho_{\gamma})$ denotes the ratio between the energy density of the baryons and the one of the photons; the shear viscosity coefficient. Finally the equations for the CDM fluctuations are simply 
\begin{equation}
\theta_{c}^{\prime} +{\mathcal H} \theta_{c} = k^2 \Phi, \qquad \delta_{c}^{\prime} = 3 \Psi^{\prime} - \theta_{c}.
\label{CDM1}
\end{equation}

We are looking for a solution of the system formed by Eqs. (\ref{evolR}), (\ref{nu1})--(\ref{nu3}), (\ref{pb2})--(\ref{pb3}) and (\ref{CDM1})  in the region $k \tau \ll k \tau_{1} \ll 1$ (and $\tau < \tau_{1}$) that corresponds to wavelengths 
larger than the Hubble radius prior to equality. 
The Bardeen potentials become, in this double limit $\Psi(k,x) = \Psi_{*}(k) x^{2 q}$ and $\Phi(k,x) = \Phi_{*}(k) x^{2 q}$ and the corresponding amplitudes are 
related ${\mathcal V}_{*}(k)$ as:
\begin{equation}
 q \, 2^{2 q + 2} [ \Phi_{*}(k) + 2 (q+1) \Psi_{*}(k)] + 9 {\mathcal V}_{*}(k) =0.
 \label{VM1}
\end{equation}
Furthermore, the direct integration of Eqs. (\ref{nu1}), (\ref{pb2}) and (\ref{CDM1}) leads to the following results:
\begin{equation}
\delta_{\gamma}(k,x) = \delta_{\nu}(k,x) \simeq \biggl[ 4 \Psi_{*}(k) + \frac{9}{q \, 2^{2 q+1}} {\mathcal V}_{*}(k) \biggr] x^{2 q},\qquad 
\delta_{c}(k,x) = \delta_{b}(k,x) \simeq 3  \Psi_{*}(k) x^{2 q}.
\label{VM3}
\end{equation}
Inserting Eq. (\ref{VM3}) into Eqs. (\ref{nu3}) and (\ref{pb3}) the three-divergences of the peculiar velocities can be obtained:
\begin{eqnarray}
&&\theta_{\nu}(k,x) = \frac{k^2 \tau_{1}}{(2 q +1)} \biggl[ \Psi_{*}(k) + \Phi_{*}(k) + \frac{9  ( 1- 2 q)}{q\,2^{2 q + 3} }\,{\mathcal V}_{*}(k) \biggr] x^{2 q +1},
\nonumber\\
&& \theta_{\gamma\,b}(k, x) = \frac{k^2 \tau_{1}}{(2 q +1)} \biggl[ \Psi_{*}(k) + \Phi_{*}(k) + \frac{9}{q\, 2^{2 q+3}} \biggl(1 - \frac{3 q}{2 R_{b}} \biggr) {\mathcal V}_{*}(k)\biggr] x^{2 q +1},
\label{VM5}
\nonumber\\
&&  \theta_{c}(k,x) = \frac{k^2 \tau_{1}}{2 (q+1)} \,\Phi_{*}(k) \, x^{2 q+1}.
\end{eqnarray}
Finally direct integration of Eq. (\ref{nu3}) implies an explicit expression for the dimensionless anisotropic stress of the neutrinos:
\begin{equation}
\sigma_{\nu}(k,x) = \frac{2 k^2 \tau_{1}^2  [ \Psi_{*}(k)  + \Phi_{*}(k)] }{15 ( q +1) ( 2 q+1) }x^{2(q +1)} + \frac{3 k^2 \tau_{1}^2}{5 \,q\,2^{2 q+2}} \biggl(\frac{1- 2q}{1+2 q} \biggr) {\mathcal V}_{*}(k) \, x^{2 (q+1)},
\label{anSTR1}
\end{equation}
where the hierarchy of the neutrinos has been truncated by requiring ${\mathcal F}_{\nu 3}=0$, as it happens in the case of the adiabatic mode and of some of the entropic solutions. Equation (\ref{anSTR1}) must also be compatible 
with the $(i\neq j)$ component of the perturbed Einstein equation, i.e. $\nabla^4 (\Phi - \Psi) = 12 \pi G a^2 \partial_{i} \partial_{j} \Pi^{ij}$;  in the concordance paradigm  the total anisotropic stress coincides with the neutrino component and it is given by $\partial_{i} \partial_{j} \Pi^{ij} = 4\rho_{\nu} \nabla^2 \sigma_{\nu}/3$. This relation generates a further condition on $\sigma_{\nu}$ i.e.
$6 R_{\nu} \sigma_{\nu}(k,x) = k^2 \tau_{1}^2 ( \Psi_{*} - \Phi_{*}) x^{2 (q +1)}$ where $R_{\nu} = \rho_{\nu}/(\rho_{\gamma}+ \rho_{\nu})$ denotes the neutrino fraction in the radiation plasma. Using this expression for $\sigma_{\nu}$ together with Eq. (\ref{anSTR1}) it is easy to obtain an algebraic relation between $\Phi_{*}(k)$, 
$\Psi_{*}(k)$ and ${\mathcal V}_{*}(k)$; such a relation can be inserted back into Eq.  (\ref{VM1}) to derive  $\Phi_{*}(k)$ and $\Psi_{*}(k)$ in terms of ${\mathcal V}_{*}(k)$:
\begin{eqnarray}
\Phi_{*}(k) &=& \frac{9 {\mathcal V}_{*}(k)}{2^{2 (q+1)}\, q} \,\frac{[q\,(4 R_{\nu} -5) - 5]}{ [4 R_{\nu} + 5 (q+1) (2 q + 3)]},
\nonumber\\
\Psi_{*}(k)  &=& -\frac{9 {\mathcal V}_{*}(k)}{2^{2(q+1)}\, q} \,\frac{[2 R_{\nu} + 5 ( q+1)]}{ [4 R_{\nu} + 5 ( 2 q+3) (q+1)]}.
\label{anSTR4}
\end{eqnarray}
For the sake of accuracy it is finally appropriate to mention that the solution derived here satisfies the Hamiltonian and the momentum 
constraints stemming fro the $(00)$ and $(0i)$ components of the perturbed Einstein equations. 

The analytic form of the viscous mode for typical wavelengths larger than the Hubble radius prior to matter-radiation 
equality can be used as initial condition of the Einstein-Boltzmann hierarchy.  To illustrate this point and to spell out more clearly this strategy the 
power spectra of the adiabatic mode, of the viscous mode and of their cross-correlation can be assigned at the common pivot scale $k_{\mathrm{p}} =0.002\,\,\mathrm{Mpc}^{-1}$ which corresponds to $\ell_{\mathrm{p}} \simeq 30$ in multipole space. The power spectra will be denoted by the unified notation  
$P_{X}(k)= {\mathcal A}_{X} (k/k_{p})^{n_{y}-1}$ where $X= {\mathcal R},\, {\mathcal V},\, {\mathcal R}{\mathcal V}$ are, respectively the amplitudes of the adiabatic mode, of the viscous mode and of their mutual correlation; similarly $y= r,\,v,\,rv$ denote the corresponding spectral indices. 
The temperature anisotropies for multipoles $\ell < \sqrt{z_{{\mathrm{rec}}}}$ (where $z_{{\mathrm{rec}}} \simeq 1090.51$ is the redshift of recombination) are unaffected by the thickness of the visibility function \cite{visibility}. The SW contribution typically 
peaks for comoving wavenumbers $k \simeq 0.0002 \, \mathrm{Mpc}^{-1}$ while the integrated Sachs-Wolfe effects contributes between $k_{\mathrm{min}} =0.001\, \mathrm{Mpc}^{-1}$ and $k_{\mathrm{max}}= 0.01\, \mathrm{Mpc}^{-1}$. Recalling that $\delta_{\gamma} = 4 \Delta_{I0}$ where $\Delta_{I0}$ is the monopole of the intensity of the radiation field, from Eq. (\ref{pb2}) we have that 
\begin{equation}
\frac{\delta_{\gamma}(k, \tau_{\mathrm{rec}}) - \delta_{\gamma}(k, \tau_{i})}{4}= \Psi(k,\tau_{\mathrm{rec}}) -\Psi(k,\tau_{i}) + \frac{9}{4} {\mathcal V}_{*}(k) \int_{\alpha_{i}}^{\alpha_{\mathrm{rec}}} d\beta \frac{\beta^{2 q -1}}{(3 \beta + 4)^{2 q}},
\label{SW1}
\end{equation}
where $\tau_{i}$ corresponds to the initial time at which the normalization is set (typically $\tau_{i} \ll \tau_{\mathrm{eq}}$ and $\alpha_{i} \ll 1$). The SW contribution is then given by:
\begin{eqnarray}
\Delta^{(SW)}_{I}(k,\mu,\tau_{0}) &=& - \biggl[\frac{{\mathcal R}_{*}(k)}{5} + f(q) {\mathcal V}_{*}(k)\biggr] e^{ - i \mu k\tau_{0}}, 
\nonumber\\
f(q) &=&  \frac{3^{ 3 - 2q}}{20 q} \biggl\{ 1 - \frac{5}{6} \biggl( \frac{3 \alpha_{\mathrm{rec}}}{4} \biggr)^{2 q} F[ 2q,\, 2q,\, 1+ 2 q,\, - 
3 \alpha_{\mathrm{rec}}/4] \biggr\},
\label{SWcont}
\end{eqnarray}
where $F[a, b, c, z]$ denotes the hypergeometric function and $\mu= \hat{k}\cdot\hat{n}$ is the projection of the direction of the Fourier mode in the direction of the photon momentum\footnote{In the limit $\alpha_{\mathrm{rec}} \to 0$ the second term in the bracket goes to zero. 
In the limit $\alpha_{\mathrm{rec}} \gg 1$ the same term provides a logarithmic correction. To simplify Eq. (\ref{SWcont})  the limit $\alpha_{i} = 0$ 
has been assumed.}. Following a standard procedure and denoting $k_{0} = \tau_{0}^{-1}$ the angular power spectrum of the full SW contribution will be given by: 
\begin{eqnarray}
&& C_{\ell}^{(SW)} = \frac{{\mathcal A}_{{\mathcal R}}}{25} {\mathcal Z}_{{\mathcal R}}(n_{r},\ell) 
+ f^2(q){\mathcal A}_{{\mathcal V}} {\mathcal Z}_{{\mathcal V}}(n_{v},\ell) 
+ \frac{2f(q)}{5} {\mathcal A}_{{\mathcal R}{\mathcal V}} {\mathcal Z}_{{\mathcal R}{\mathcal V}}(n_{rv},\ell) \cos{\gamma_{rv}},
\label{SCPS}\\
&& {\mathcal  Z}_{X}(n_{x},\ell) =\frac{\pi^2}{4} \biggl(\frac{k_{0}}{k_{\rm p}}\biggr)^{n_{x}-1}\,\,2^{n_{x}}\frac{\Gamma( 3 - n_{x}) 
\Gamma\biggl( \ell + \frac{n_{x}}{2} - 
\frac{1}{2}\biggr)}{\Gamma^2\biggl(\frac{4 - n_{x}}{2}\biggr)
 \Gamma\biggl( \frac{5}{2} + \ell - \frac{n_{x}}{2} \biggr) },
\label{SW2}
\end{eqnarray}
where $\gamma_{rv}$ denotes the correlation angle. Equations (\ref{SCPS}) and (\ref{SW2}) can be used as starting point for dedicated analyses where 
the dominant adiabatic mode is correlated (or anticorrelated) with the viscous mode.

The viscous curvature perturbations vanish faster than in the case of the conventional entropic modes.
The baryon-radiation and the CDM-radiation modes lead to Bardeen potentials and curvature perturbations that  vanish as $\alpha$ in the 
limit $\alpha \ll 1$; in the same limit  the curvature perturbations induced by the viscous mode vanish as $\alpha^{2 q}$ with $q \geq1 $. The curvature perturbations corresponding to the neutrino entropy mode  are identically zero (to leading order) while the Bardeen potentials are approximately constant. The neutrino velocity mode   arises when the momentum constraint is vanishing at early times and, simultaneously, the total density is uniform. This requirement, once inserted into the evolution equations in the longitudinal gauge, entails a singularity in the Bardeen potentials; this mode (singular in the longitudinal description) is regular in the synchronous gauge
and has no direct counterpart in the viscous case. These considerations show that prior to matter-radiation equality the viscous solution is really and truly of isocurvature type since, in this limit, the curvature perturbations tend to zero faster than in the case of their (regular) entropic counterparts.

All in all the results of this investigation demonstrate the existence  of a regular class of viscous solutions for the fluctuations modes of the predecoupling plasma. 
Since the CMB anisotropies are sensitive not only to the homogeneous matter content of the Universe but also to the Cauchy data of the cosmological perturbations,  
the viscous modes could serve as sound initial conditions of the Einstein Boltzmann hierarchy in conjunction with a dominant adiabatic component. In the present framework the direct observational limits stemming from the temperature and polarization 
anisotropies of the CMB  can be translated into specific constraints on the properties of the bulk viscous stresses across matter-radiation equality.

\end{document}